\documentclass[pra,aps,twocolumn, nobalancelastpage, superscriptaddress]{revtex4-2}
\usepackage[]{natbib}
\usepackage{graphicx}% Include figure files
\usepackage{dcolumn}% Align table columns on decimal point
\usepackage{bm}% bold math
\usepackage[separate-uncertainty=true]{siunitx}
\usepackage{layouts}
\usepackage{epstopdf}
\usepackage{bbm}
\usepackage{bm}
\usepackage{hyperref}
\begin{document}

%\preprint{AIP/123-QED}

\title[]{High emission rate from a Purcell-enhanced, triggered source of pure single photons in the telecom C-band}% Force line breaks with \\

\author{Cornelius Nawrath}
\email{c.nawrath@ihfg.uni-stuttgart.de}
\altaffiliation{These authors contributed equally to this work}
\affiliation{Institut f\"ur Halbleiteroptik und Funktionelle Grenzfl\"achen, \\Center for Integrated Quantum Science and Technology (IQ\textsuperscript{ST}) and SCoPE, \\University of Stuttgart, Allmandring 3, 70569 Stuttgart, Germany}

\author{Raphael Joos}
\altaffiliation{These authors contributed equally to this work}
\affiliation{Institut f\"ur Halbleiteroptik und Funktionelle Grenzfl\"achen, \\Center for Integrated Quantum Science and Technology (IQ\textsuperscript{ST}) and SCoPE, \\University of Stuttgart, Allmandring 3, 70569 Stuttgart, Germany}

\author{Sascha Kolatschek}\affiliation{Institut f\"ur Halbleiteroptik und Funktionelle Grenzfl\"achen, \\Center for Integrated Quantum Science and Technology (IQ\textsuperscript{ST}) and SCoPE, \\University of Stuttgart, Allmandring 3, 70569 Stuttgart, Germany}

\author{Stephanie Bauer}\affiliation{Institut f\"ur Halbleiteroptik und Funktionelle Grenzfl\"achen, \\Center for Integrated Quantum Science and Technology (IQ\textsuperscript{ST}) and SCoPE, \\University of Stuttgart, Allmandring 3, 70569 Stuttgart, Germany}

\author{Pascal Pruy}\affiliation{Institut f\"ur Halbleiteroptik und Funktionelle Grenzfl\"achen, \\Center for Integrated Quantum Science and Technology (IQ\textsuperscript{ST}) and SCoPE, \\University of Stuttgart, Allmandring 3, 70569 Stuttgart, Germany}

\author{Florian Hornung}\affiliation{Institut f\"ur Halbleiteroptik und Funktionelle Grenzfl\"achen, \\Center for Integrated Quantum Science and Technology (IQ\textsuperscript{ST}) and SCoPE, \\University of Stuttgart, Allmandring 3, 70569 Stuttgart, Germany}

\author{Julius Fischer}\affiliation{Institut f\"ur Halbleiteroptik und Funktionelle Grenzfl\"achen, \\Center for Integrated Quantum Science and Technology (IQ\textsuperscript{ST}) and SCoPE, \\University of Stuttgart, Allmandring 3, 70569 Stuttgart, Germany}
\affiliation{Current Affiliation: QuTech and Kavli Institute of Nanoscience, Delft University of Technology, Delft 2628 CJ, Netherlands}

\author{Jiasheng Huang}\affiliation{Institut f\"ur Halbleiteroptik und Funktionelle Grenzfl\"achen, \\Center for Integrated Quantum Science and Technology (IQ\textsuperscript{ST}) and SCoPE, \\University of Stuttgart, Allmandring 3, 70569 Stuttgart, Germany}

\author{Ponraj Vijayan}\affiliation{Institut f\"ur Halbleiteroptik und Funktionelle Grenzfl\"achen, \\Center for Integrated Quantum Science and Technology (IQ\textsuperscript{ST}) and SCoPE, \\University of Stuttgart, Allmandring 3, 70569 Stuttgart, Germany}

\author{Robert Sittig}\affiliation{Institut f\"ur Halbleiteroptik und Funktionelle Grenzfl\"achen, \\Center for Integrated Quantum Science and Technology (IQ\textsuperscript{ST}) and SCoPE, \\University of Stuttgart, Allmandring 3, 70569 Stuttgart, Germany}

\author{Michael Jetter}\affiliation{Institut f\"ur Halbleiteroptik und Funktionelle Grenzfl\"achen, \\Center for Integrated Quantum Science and Technology (IQ\textsuperscript{ST}) and SCoPE, \\University of Stuttgart, Allmandring 3, 70569 Stuttgart, Germany}

\author{Simone Luca Portalupi}\affiliation{Institut f\"ur Halbleiteroptik und Funktionelle Grenzfl\"achen, \\Center for Integrated Quantum Science and Technology (IQ\textsuperscript{ST}) and SCoPE, \\University of Stuttgart, Allmandring 3, 70569 Stuttgart, Germany}

\author{Peter Michler}\affiliation{Institut f\"ur Halbleiteroptik und Funktionelle Grenzfl\"achen, \\Center for Integrated Quantum Science and Technology (IQ\textsuperscript{ST}) and SCoPE, \\University of Stuttgart, Allmandring 3, 70569 Stuttgart, Germany}

\date{\today}

\begin{abstract}
Several emission features mark semiconductor quantum dots as promising non-classical light sources for prospective quantum implementations. For long-distance transmission~\cite{Liao2017} and Si-based on-chip processing\cite{Qiang2018,Bauer2021}, the possibility to match the telecom C-band~\cite{Cao2019} stands out, while source brightness and high single-photon purity are key features in virtually any quantum implementation~\cite{Takemoto,Bozzio2022}. Here we present an InAs/InGaAs/GaAs quantum dot emitting in the telecom C-band coupled to a circular Bragg grating. The Purcell enhancement of the emission enables a simultaneously high brightness with a fiber-coupled single-photon count rate of \SI{13.9}{\mega\hertz} for an excitation repetition rate of \SI{228}{\mega\hertz} (first-lens collection efficiency $\sim$\SI{17}{\percent} for $\mathrm{NA}=0.6$), while maintaining a low multi-photon contribution of $g^{(2)}(0)=0.0052$. Moreover, the compatibility with temperatures of up to \SI{40}{\kelvin} attainable with compact cryo coolers, further underlines the suitability for out-of-the-lab implementations.
\end{abstract}

\keywords{Suggested keywords}
\maketitle
Efficient and on-demand sources of single photons have a number of applications ranging from  quantum communication~\cite{Takemoto,Xu2020,Bozzio2022} and cryptography~\cite{Xu2020,Bozzio2022} to quantum metrology~\cite{Georgieva2021}. An emission wavelength in the telecom C-band (\SI{1530}{\nano\meter} to \SI{1565}{\nano\meter}) is especially sought-after due to the compatibility with the mature silicon photonics platform~\cite{Qiang2018,Bauer2021}, as well as for both free space (low solar background and Rayleigh scattering)~\cite{Liao2017}, and fiber-based transmission (absorption minimum and low dispersion).\\
Among non-classical light emitters around \SI{1550}{\nano\meter}~\cite{Cao2019}, semiconductor quantum dots (QDs) on InP have been investigated for two decades~\cite{Takemoto2007,Birowsuto,Takemoto,Miyazawa2016,Kors2017,Muller2018,Shooter2020,Anderson2020a,Musia2021,Holewa2021}, culminating in application-oriented implementations such as quantum key distribution over \SI{120}{\kilo\meter} of fiber~\cite{Takemoto}, teleportation of time-bin qubits~\cite{Anderson2020a} and distribution of single photons, and entangled photon pairs over \SI{4.6}{\kilo\meter} of deployed fiber~\cite{Shooter2020}. More recently, InAs QDs on GaAs basis have gained increasing interest, as the material system is widespread in research and industry due to its low cost and well-developed processing techniques. To alleviate the strain between InAs and GaAs, which limits the emission of the current state-of-the-art QDs to the near-infrared regime, metamorphic buffer layers (MMBs)~\cite{Semanova,Paul2017,Sittig2022} have been employed, enabling operation in the telecom C-band and opening the route for a number of studies mainly on the fundamental properties for quantum applications, i.e. single-photon emission~\cite{Paul2017,Nawrath2019,Zeuner2021,Nawrath2021}, entangled photon pair emission~\cite{Olbr,Zeuner2021,Lettner2021} and photon indistinguishability~\cite{Nawrath2019,Nawrath2021}.\\
While the feasibility of device integration as an advantage of the solid-state environment has been exploited around \SI{1550}{\nano\meter}~\cite{Muller2018,Shooter2020}, few studies report on tackling the related brightness limitation due to total internal reflection. For the InP material system mesas~\cite{Musia2021,Holewa2021}, optical horn structures~\cite{Takemoto2007,Takemoto, Miyazawa2016} and photonic crystal cavities~\cite{Birowsuto,Kors2017} have been reported, claiming up to \SI{13.3}{\percent} of collection efficiency into the first lens (NA$=0.4$)~\cite{Musia2021}. For GaAs-based QDs, the recently reported efficient MMB design~\cite{Sittig2022} enabling the realization of $\lambda$-cavities in growth direction, has opened the possibility for the incorporation into high-quality photonic structures~\cite{Sittig2022,Bremer2022}.\\
Among the possible implementations of such cavities, the circular Bragg grating (CBG)~\cite{Liu2019,Wang2019} offers the simultaneous advantage of Purcell enhancement of the radiative decay, as well as a broadband increase of the collection efficiency. With this, CBGs are capable of simultaneously supporting both transitions of the biexciton-exciton cascade, generating polarization-entangled photon pairs~\cite{Liu2019,Wang2019}. CBGs have been reported for the telecom O-band~\cite{Kolatschek2021,Xu2022,Barbiero2022a} and the design parameters have been investigated by simulations for the GaAs and InP system~\cite{Bremer2022,Barbiero2022} for the telecom C-band, however, an experimental realization had yet to be demonstrated.
\begin{figure}[h!]
\centering
\includegraphics[width=0.99\linewidth]{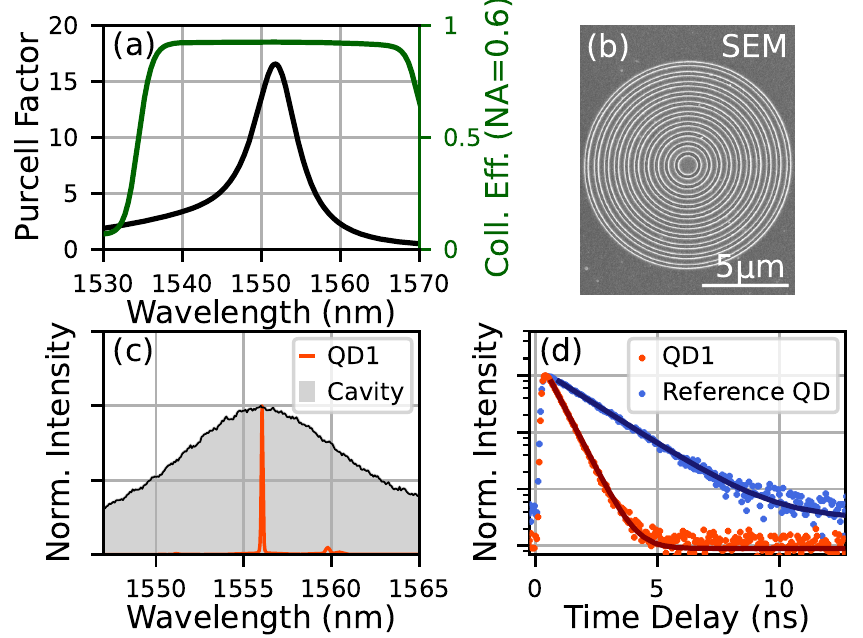}
\caption{CBG Characterization: (a) FDTD simulations for Purcell factor and collection efficiency (NA$=0.6$), (b) SEM image of the CBG, (c) Spectrum of the cavity mode under strong, non-resonant pumping, and of QD1 under pulsed p-shell excitation, (d) TCSPC measurements with fits (solid lines) under non-resonant excitation at \SI{850}{\nano\meter} of QD1 (decay time \SI{0.52}{\nano\second}) and a reference QD (decay time \SI{1.59}{\nano\second}) in a semi-logarithmic scale.}
\label{fig:fig1}
\end{figure}

Here, we close this gap, investigating an InAs/InGaAs/GaAs QD emitting in the telecom C-band incorporated into a non-deterministically placed CBG, reaching a Purcell enhancement of a factor of 3. The device combines a simultaneously high first-lens collection efficiency of \SI{17.4}{\percent} ($\mathrm{NA}=0.6$) and a fiber-coupled single-photon count rate (FCSPCR) of \SI{4.77e6}{\per\second} at \SI{76}{\mega\hertz} repetition rate (end-to-end-efficiency of \SI{6.3}{\percent}), with a low multi-photon contribution $\left(g^{(2)}_{\mathrm{analyt}}(0)=\SI{7.2(40)E-3}{}\right)$. We further demonstrate the thermal stability of these properties up to temperatures attainable with compact cryocoolers: at \SI{40}{\kelvin} a count rate of \SI{2.96e6}{\per\second} is achieved with $g^{(2)}_{\mathrm{analyt}}(0)=\SI{4.32(13)E-2}{}$. Repetition rates up to \SI{228}{\mega\hertz} are investigated, showing that the Purcell factor of 3 allows to maintain high brightness and purity values ($\mathrm{FCSPCR=\SI{13.88E6}{\per\second}}$, $g^{(2)}_{\mathrm{analyt}}(0)=\SI{5.2(1)E-3}{}$ at \SI{228}{\mega\hertz}). To the best of our knowledge, this FCSPCR exceeds highest value reported for a QD-based single-photon source in the telecom C-band by around two orders of magnitude~\cite{Musia2021}, while more than a factor of 4 is found including results in the telecom L-band~\cite{Takemoto}. Finally, we measure the two-photon interference (TPI) visibility of the emission.\\

\begin{figure*}[t]
\centering
\includegraphics[width=0.99\textwidth]{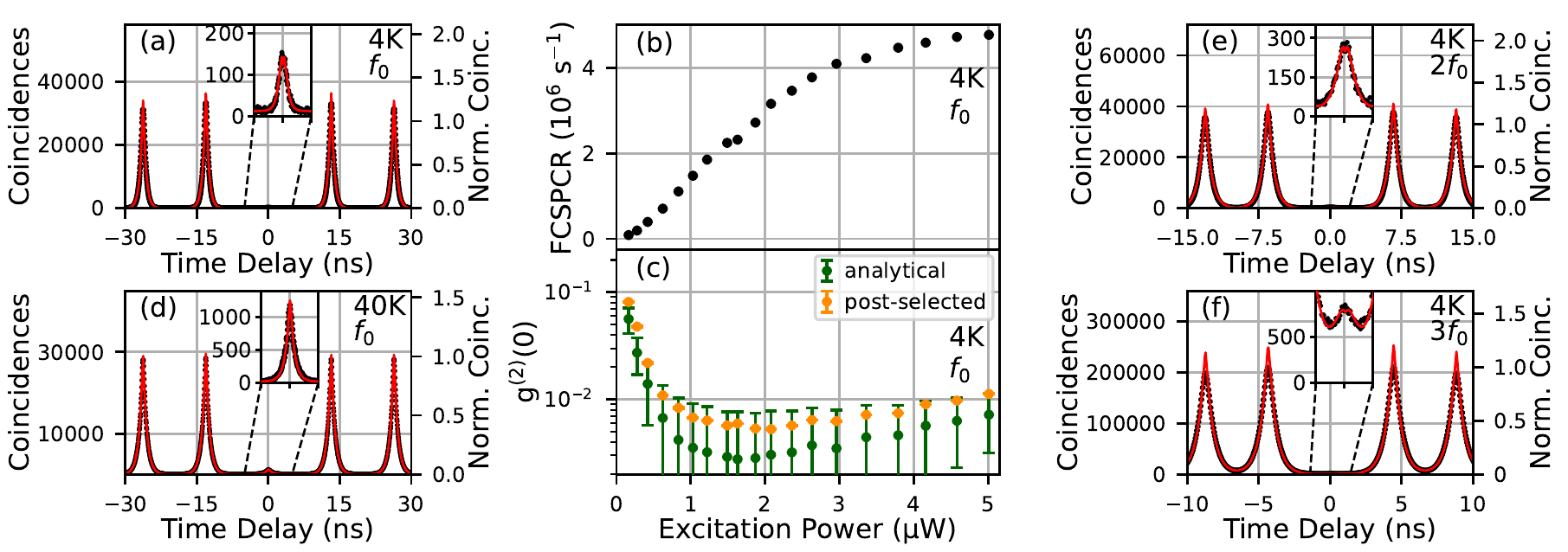}
\caption{Brightness and single-photon purity of QD1: (a) $g^{(2)}(\tau)$ at \SI{4}{\kelvin}, $P=\SI{5.01}{\micro\watt}$ and $f_{0}=\SI{76}{\mega\hertz}$, (b) and (c) fiber-coupled single-photon count rate (FCSPCR) after the full setup at the detector, corrected for the detection efficiency, and $g^{(2)}(0)$ over the excitation power, (d) $g^{(2)}(\tau)$ at \SI{40}{\kelvin} and $P=\SI{5.01}{\micro\watt}$, (e) and (f) $g^{(2)}(\tau)$ at \SI{4}{\kelvin}, $P=\SI{10.02}{\micro\watt}$, $2f_{0}=\SI{152}{\mega\hertz}$ and $P=\SI{15.03}{\micro\watt}$, $3f_{0}=\SI{228}{\mega\hertz}$, respectively.}
\label{fig:fig2}
\end{figure*}

The optimal performance of the device is investigated via finite-difference time-domain (FDTD) simulations as presented in Fig.~\ref{fig:fig1}(a). The collection efficiency (NA$=0.6$) is above \SI{85}{\percent} for almost the entire telecom C-band, with maximum values of more than \SI{92}{\percent}. Geometrical brightness enhancement in combination with the maximum Purcell factor of 16.6 (FWHM$=\SI{7.45}{\nano\meter}$) makes this device design appealing for bright emission of single photons and entangled photon pairs~\cite{Liu2019,Wang2019}.\\ 
A scanning electron microscopy (SEM) image of the finished device can be seen in Fig.~\ref{fig:fig1}(b). The cavity mode, taken as the spectrum under strong, non-resonant excitation (shaded curve in Fig.~\ref{fig:fig1}(c)) exhibits a FWHM of \SI{9.74\pm1.8}{\nano\meter} and is in good agreement with the mode expected from the FDTD simulations. The slightly larger width is in part due to a minor ellipticity, leading to a polarization mode splitting of \SI{1.24\pm0.03}{\nano\meter}.\\
The investigated QD transition, abbreviated as QD1 from here on, (solid orange line in Fig.~\ref{fig:fig1}(c)), in resonance with the cavity mode is excited via its p-shell $\sim$\SI{1530}{\nano\meter}). We attribute this transition to a positively charged trion (see supplementary material). Note that other QD candidates with similar properties are found on the same sample (see supplementary).\\
%To assess a possible Purcell enhancement of QD1, we take time-correlated single-photon counting (TCSPC) measurements of the decay time of this, as well as 6 other exemplary QD trion transitions on the same sample outside of CBG structures. For consistency all transitions are excited non-resonantly at \SI{875}{\nano\meter}. Fig.~\ref{fig:fig1}(d) shows the decay traces alongside the fit functions (solid lines) of QD1 (orange, decay time \SI{0.52}{\nano\second}), as well as a reference QD transition (blue, decay time \SI{1.59}{\nano\second}) close to the average of \SI{1.55}{\nano\second} (standard deviation \SI{0.37}{\nano\second}) of the reference QDs. This results in a Purcell factor of 3. The discrepancy between the maximally expected and the measured value is attributed mainly to a spatial offset between the center of the non-deterministically positioned CBG and the QD (see supplementary). Deterministic fabrication methods~\cite{Kolatschek2019,Liu2018,Liu2019} promise to bring the experimental value close to its optimum, benefitting the attainable transmission rates in quantum communication applications~\cite{Shooter2020}.
Time-correlated single-photon counting (TCSPC) measurements of the decay time as displayed in Fig.~\ref{fig:fig1}(d), yield a value of $\SI{0.52}{\nano\second}$ for QD1 and a mean value of \SI{1.55}{\nano\second} (standard deviation \SI{0.37}{\nano\second}) for 6 exemplary trion transitions of QDs on the same sample outside of CBG structures. This results in a Purcell factor of 3. The discrepancy between the maximally expected and the measured value is attributed mainly to a spatial offset between the center of the non-deterministically positioned CBG and the QD (see supplementary). Deterministic fabrication methods~\cite{Kolatschek2019,Liu2018,Liu2019} promise to bring the experimental value close to its optimum, benefitting the attainable transmission rates in quantum communication applications~\cite{Shooter2020}.

Key figures of merit of such schemes are the brightness and the single-photon purity~\cite{Takemoto,Bozzio2022}. The latter is evaluated via the second-order auto-correlation function $g^{(2)}(\tau)$, with the time delay $\tau$, for different excitation powers $P$. Fig.~\ref{fig:fig2}(a) shows an exemplary measurement result and fit function (solid line) at an excitation power $P=\SI{5.01}{\micro\watt}$ measured above the objective. For the normalization and evaluation, bunching processes up to the millisecond time scale are taken into account. The significantly suppressed peak at zero time delay exhibits an additional dip that is attributed to a slight refilling from the QD surrounding, e.g. via charge carrier trap states. The fit function takes this feature into account and upon fully analytical evaluation of the fit function excluding any offset on the ordinal, a value as low as $g^{(2)}_{\mathrm{analyt}}(0)=\SI{7.2(40)E-3}{}$ is found. As for applications any coincidences from background contributions are also relevant, the corresponding post-selected value $g^{(2)}_{\mathrm{ps}}(0)=\SI{1.13E-2}{}$ is evaluated in the full central repetition period of $\pm$\SI{6.58}{\nano\second}, corrected for coincidences from detector dark counts (see supplementary material for details on purity evaluations and a discussion of the influence of the post-selection window on $g^{(2)}_{\mathrm{ps}}(0)$).\\
Notably, we find only a minor trade-off between this high single-photon purity and the brightness: Figs.~\ref{fig:fig2}(b) and (c) display the FCSPCR at the detector, including corrections for dark counts, multi-photon contribution and detection efficiency, and the corresponding values for $g^{(2)}(0)$ over the excitation power. The slightly reduced single-photon purities for low excitation powers are attributed to a second QD in the spatial and spectral collection area being partially excited (see supplementary). For $P=\SI{5.01}{\micro\watt}$, a value of $\mathrm{FCSPCR}=\SI{4.77e6}{\per\second}$ (raw, detected count rate \SI{3.91e6}{\per\second}) is found (see supplementary material for details). This corresponds to an end-to-end efficiency of \SI{6.3}{\percent} and a collection efficiency of \SI{17.4}{\percent} into the first optical element (NA$=0.6$), taking into account the setup efficiency (see supplementary). Note that due to the observed blinking, the source is active $\sim$\SI{70}{\percent} of the time at saturation (see supplementary material). Integrating the QD in a diode structure is expected to eliminate this limitation of the brightness~\cite{Zhai2020,Schimpf2021}. However, also in the current state the presented device outperforms all QD emitters with emission around the C-band, reported so far: even with sophisticated structures on the well-established InP material basis~\cite{Takemoto2007,Takemoto,Miyazawa2016,Musia2021,Holewa2021}, \SI{13.3}{\percent} of first-lens collection efficiency~\cite{Holewa2021} have not been surpassed. Apart from this purely source-related brightness measure, the efficiency of the collection and filtering determines the usable count rate in applications, but has rarely been reported with high efficiency~\cite{Takemoto2007,Takemoto,Miyazawa2016} (record end-to-end efficiency so far in the telecom L-band: \SI{5}{\percent}~\cite{Takemoto}). The presented work thus signifies an important advance towards out-of-the-lab implementations.\\
Moreover, the device is stable at elevated temperatures: the measurement of $g^{(2)}(\tau)$ at a temperature of \SI{40}{\kelvin} and $P=\SI{5.01}{\micro\watt}$, as displayed in Fig.~\ref{fig:fig2}(d), results in values of $g^{(2)}_{\mathrm{analyt}}(0)=\SI{4.32(13)e-2}{}$ and $g^{(2)}_{\mathrm{ps}}(0)=\SI{4.89e-2}{}$. A maximum of $\mathrm{FCSPCR}=\SI{2.96e6}{\per\second}$ (raw, detected count rate \SI{2.47e6}{\per\second}) is found. This temperature stability allows the implementation of the source in compact, economic Stirling cryocoolers.\\
A further decisive advantage is the Purcell-enhanced decay time. It allows for higher excitation repetition rates, while maintaining high single-photon purity. Fig.~\ref{fig:fig2}(e) and (f) display measurements of $g^{(2)}(\tau)$ at \SI{152}{\mega\hertz} and \SI{228}{\mega\hertz} repetition rate, \SI{4}{\kelvin} and the same excitation power per pulse as the measurements displayed in Fig.~\ref{fig:fig2}(a) and (b). Values of $g^{(2)}_{\mathrm{analyt}}=\SI{1.08(56)E-2}{}$ ($g^{(2)}_{\mathrm{ps}}=\SI{1.51E-2}{}$) and $g^{(2)}_{\mathrm{analyt}}=\SI{5.2(1)E-3}{}$ ($g^{(2)}_{\mathrm{ps}}=\SI{2.32E-2}{}$) are found, respectively (see supplementary for further measurements). The corresponding maximum values of FCSPCR=\SI{9.44e6}{\per\second}, raw, detected countrate \SI{7.75e6}{\per\second}) and FCSPCR=\SI{13.88e6}{\per\second} (raw, detected countrate \SI{11.37e6}{\per\second}), respectively, further underline the suitability of the presented source for quantum applications. Note that especially at high repetition rates, due to the overlap of the peaks, a slight reduction of the post-selection window further decreases $g^{(2)}_{\mathrm{ps}}(0)$ while maintaining a high count rate (see supplementary).

%To further assess the coherence of the emitted photons, the lineshape is acquired by Fourier-transform spectroscopy measuring the first-order auto-correlation function $g^{(1)}(\tau)$. The visibility of the interference fringes as a function of the temporal delay is displayed in Fig.~\ref{fig:fig3}(a) exemplarily for an excitation power of \SI{3.8}{\micro\watt}. The fit function is the Fourier-transform of the Voigt lineshape (subscript V) of the emission on the investigated long time scales of seconds. The resulting FWHM$_{\mathrm{V}}$ for different excitation powers is plotted in Fig.~\ref{fig:fig3}(b). In agreement with previous studies~\cite{Nawrath2019,Nawrath2021,Sittig2022} we find predominantly inhomogeneously broadened lineshapes with a mean value of FWHM$_\mathrm{V}$ of \SI{15.8}{\giga\hertz} (standard deviation \SI{0.4}{\giga\hertz}) and little excitation power dependence. Instabilities of the magnetic~\cite{Kuhlmann2015,Malein2016} or electric field~\cite{Kuhlmann2015,Reigue2019} at the site of the QD are assumed to be responsible for the decoherence of the emission. Ref.~\cite{Liu2018} suggests that the proximity to surface states should not affect the emission negatively for the device dimensions. Such that active stabilization of the magnetic~\cite{Kuhlmann2015,Malein2016} and electric field~\cite{Kuhlmann2015,Reigue2019}, is expected to improve the coherence more than the applciation of a passivation layer~\cite{Liu2018}.\\
\begin{figure}[h!]
\centering
\includegraphics[width=0.99\linewidth]{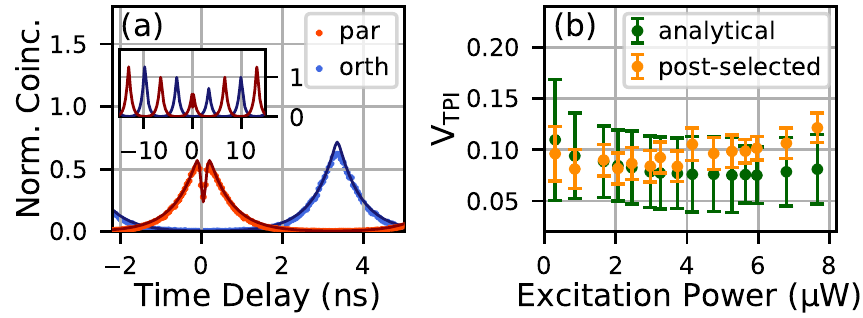}
\caption{TPI Visibility of QD1: (a) exemplary TPI measurement for parallel (par) and orthogonal (orth) polarization at $P=\SI{7.7}{\micro\watt}$. The orthogonal data and fit are offset horizontally for visual clarity. (b) $V_{\mathrm{TPI}}$ for different excitation powers.}
\label{fig:fig3}
\end{figure}
For a large number of implementations in quantum information and cryptography or QKD~\cite{Takemoto,Bozzio2022}, high brightness and purity are the only requirements. Nonetheless, here we benchmark the photon coherence which is of importance in other applications. To this end, the degree of indistinguishability (TPI visibility, $V_{\mathrm{TPI}}$) of photons emitted \SI{6.6}{\nano\second} apart (excitation repetition rate \SI{152}{\mega\hertz}) is probed. An exemplary correlation histogram at $P=\SI{7.7}{\micro\watt}$ for the measurements with parallel and orthogonal (completely distinguishable) polarization is displayed in Fig.~\ref{fig:fig3}(a) alongside the respective fit functions. The partial suppression of the central peak for parallel polarization results in a value of $V_{\mathrm{TPI,analyt}}=\SI{8.10\pm3.38}{\percent}$. Similarly to the linewidth (see supplementary), $V_{\mathrm{TPI}}$ exhibits little dependence on the excitation power, as is shown in Fig.~\ref{fig:fig3}(b) for the values via analytical and post-selected evaluation. Despite the Purcell enhancement of the decay rate, these results are comparable to $V_{\mathrm{TPI}}$ found under purely resonant excitation for a QD with only a bottom DBR~\cite{Nawrath2021}. Note, however, that in the latter the signal was filtered to \SI{5}{\giga\hertz} (see supplementary material for further measurements and discussions on $V_{\mathrm{TPI}}$). Instabilities of the magnetic~\cite{Kuhlmann2015,Malein2016} or electric field~\cite{Kuhlmann2015,Reigue2019} at the site of the QD are assumed to be responsible for the decoherence of the emission. Ref.~\cite{Liu2018} suggests that the proximity to the surface should not affect the emission negatively for the device dimensions. Thus, the stabilization of the magnetic~\cite{Kuhlmann2015,Malein2016} and electric field~\cite{Kuhlmann2015,Reigue2019}, is expected to improve the coherence more than the application of a passivation layer~\cite{Liu2018}. Such implementations promise to combine the high brightness and purity of the emission with a high degree of photon indistinguishability and coherence, further broadening the range of applications for which the source is suitable.

In summary, a bright source of single photons emitting in the telecom C-band based on an InAs/InGaAs/GaAs QD under p-shell excitation in a circular Bragg grating is presented. A simultaneously high fiber-coupled single-photon count rate of \SI{13.88}{\mega\hertz} for an excitation repetition rate of \SI{228}{\mega\hertz} (first-lens collection efficiency $\sim$\SI{17}{\percent} for $\mathrm{NA}=0.6$) is demonstrated, while maintaining a low multi-photon contribution of $g^{(2)}_{\mathrm{analyt}}(0)=\SI{5.2(1)E-3}{}$ due to the Purcell factor of 3. Furthermore, the compatibility with compact cryocoolers is demonstrated, opening routes for real-world implementations. Finally, the degree of indistinguishability is investigated, identifying possible strategies to improve the coherence of the emission.

\medskip

\noindent\textbf{Methods}
The sample is based on the recently reported thin MMB design and corresponding QD growth optimizations~\cite{Sittig2022}. For the device fabrication~\cite{Kolatschek2021}, a \SI{350}{\nano\meter} Al$_{2}$O$_{3}$ spacer layer and a \SI{100}{\nano\meter} Au layer are deposited on the sample. Using a flip-chip process, it is transfered to a Si carrier with Su-8 resist as glue. By means of citric acid and hydrofluoric acid, the GaAs substrate and the AlGaAs sacrificial layer are removed. Finally, the CBG rings are fabricated by electron beam lithography with AR-P 6200 resist to produce a SiO$_{2}$ hard mask, which is used to transfer the trenches into the InGaAs membrane via ICP-RIE etching.\\
A continuous-flow helium cryostat is used to cool the sample down to \SI{4}{\kelvin}. For the experiments at \SI{40}{\kelvin}, the built-in heating element is employed. A microscope objective ($\mathrm{NA}=0.6$) is used to focus the excitation laser onto the sample and collect the QD signal. The excitation and collection beam path are overlaid using a coated mirror (reflectivity \SI{98.5}{\percent}). Volume Bragg elements are used for filtering: two notch filters ($\mathrm{FWHM}=\SI{1.2}{\nano\meter}$) suppress the excitation laser signal in the detection path and a reflective bandpass filter ($\mathrm{FWHM}=\SI{0.55}{\nano\meter}$) selects the QD transition in question. Since the SNSPDs are polarization sensitive, the two polarization components of the QD signal are split and guided to two different detectors. To this end, a fiber-coupled polarizing beam splitter is employed.\\
For the optical excitation, ps pulses at repetition rates between \SI{76}{\mega\hertz} and \SI{228}{\mega\hertz} are employed. Apart from the decay time measurements, where non-resonant excitation at \SI{875}{\nano\meter} is used, the charge carriers are excited to the p-shell of the QD $\sim$\SI{1530}{\nano\meter}. Further details on the measurement methods and the setup efficiency are presented in the supplementary material.\\

\medskip

\noindent\textbf{Funding}
The authors gratefully acknowledge the funding by the German Federal Ministry of Education and Research (BMBF) via the project QR.X (No.16KISQ013) and the European Union’s Horizon 2020 research and innovation program under Grant Agreement No. 899814 (Qurope). Furthermore, this project (20FUN05 SEQUME) has received funding from the EMPIR programme co-financed by the Participating States and from the European Union’s Horizon 2020 research and innovation programme.
\medskip

\noindent\textbf{Acknowledgements}
We want to thank Philipp Flad and Monika Ubl for their help and acknowledge the companies Single Quantum and PriTel Inc. for the persistent and timely support.

\medskip

\noindent\textbf{Author Contributions}
C.N. and R. J. performed the measurements with support from P. P.. C. N. and J. F. planned, built and optimized the setup with support from R. J.. S. K., S. B., J. H., P. V. designed and fabricated the sample. C. N., R. J. analyzed the data with support from F. H.. R. S. grew the sample with supervision of M. J.. C. N., R. J. wrote the manuscript with the help of S. L. P. and P. M.. S. L. P. and P. M. supervised the project. All authors contributed to scientific discussions.

% Bibliography
%\bibliography{Bibliography_C_Band_Bullseye_Manuscript}
%\bibliography{aipsamp}% Produces the bibliography via BibTeX.


\end{document}